\UseRawInputEncoding 
\documentclass[superscriptaddress, notitlepage, reprint]{revtex4-1}
\usepackage[english]{babel}
\usepackage{amsmath,amsthm}
\usepackage{amsfonts}
\usepackage[pdfborder={0 0 0}, colorlinks=true, urlcolor=blue, linkcolor=blue, citecolor=blue]{hyperref}
\usepackage{color}
\usepackage{graphicx}
\usepackage{float}
\usepackage{subfigure}
\usepackage{lipsum}
\usepackage{epstopdf}
\usepackage{dcolumn}
\begin{document}

\title{Quantum Thermometry with a Dissipative Quantum Rabi System}
\author{Dong  Xie}\email{xiedong@mail.ustc.edu.cn}
\affiliation{College of Science, Guilin University of Aerospace Technology, Guilin, Guangxi 541004, People's Republic of China}
\author{Chunling Xu}
\affiliation{College of Science, Guilin University of Aerospace Technology, Guilin, Guangxi 541004, People's Republic of China}
\author{An Min Wang}\email{anmwang@ustc.edu.cn}
\affiliation{Department of Modern Physics, University of Science and Technology of China, Hefei, Anhui 230026, People's Republic of China}

\begin{abstract}
Dissipative quantum Rabi System, a finite-component system composed of a single two-level atom interacting with an optical cavity field mode, exhibits a quantum phase transition, which can be exploited to greatly enhance the estimation precision of unitary parameters (frequency and coupling strength). Here, using the quantum Langevin equation, standard mean field theory and adiabatic elimination, we investigate the quantum thermometry of a thermal bath surrounding the atom with quantum optical probes. With the increase of coupling strength between the atom and the cavity field, two kinds of singularities can be observed. One type of singularity is the exceptional point (EP) in the anti-parity-time (anti-$\mathcal{PT}$) symmetrical cavity field. The other type of singularity is the critical point (CP) of phase transition from the normal to superradiant phase. We show that the optimal measurement precision occurs at the CP, instead of the EP. And the direct photon detection represents an excellent proxy for the optimal measurement near the CP. In the case where the thermal bath to be tested is independent of the extra thermal bath interacting with the cavity field, the estimation precision of the temperature always increases with the coupling strength. Oppositely, if the thermal bath to be tested is in equilibrium with the extra bath interacting with the cavity field, noises that suppress the information of the temperature will be introduced when increasing the coupling strength unless it is close to the CP.

\end{abstract}
\maketitle

\section{Introduction}
Quantum criticality provides a powerful resource for quantum metrology~\cite{lab1} due to that small variations of
physical parameters may lead to dramatic changes of quantum systems around the critical point (CP). It is attracting more and more attentions. The researches on it are mainly carried out along two approaches: one is based on the time evolution induced by a Hamiltonian close to a CP\cite{lab2}; the other is based on the ground state near a quantum phase transition ~\cite{lab3,lab4,lab5,lab6,lab7,lab8,lab9,lab10,lab11} or the general nonequilibrium steady state around a dissipative phase transition~\cite{lab12,lab13}.

An arbitrarily large estimation precision could be achieved due to that the susceptibility of the
equilibrium state diverges at the CP. However it comes at the cost of infinite steady-state preparation time~\cite{lab14}.
Ref.~\cite{lab15} provided a framework for quantum sensing enhanced by critical quantum dynamics that can relax the stringent requirement for initial state preparation. However, it still requires a longer evolution time in order to get a higher precision. Hence, quantum criticality requires sufficient time resources to show its superiority.

For the dissipative phase transition systems with long time to steady states, quantum criticality provides a powerful resource for improving the measurement precision. A lot of works have been done to improve the measurement precision of unitary parameters, such as frequency, magnetic field strength\cite{lab3,lab4,lab5,lab6,lab7,lab8,lab9,lab10,lab11,lab12,lab13,lab14,lab15}. Few studies have been conducted on non-unitary parameters, especially with regard to temperature. General systems are relatively insensitive to the non-unitary parameter-temperature.
Recently, Ref.~\cite{lab16} converted the temperature variation into a magnetic field change, then used magnetic criticality to improve the sensitivity of diamond nanothermometers.

In this article, we utilize a dissipative quantum Rabi system composed of a single two-level atom and an optical cavity field to measure the temperature of thermal bath around the atom. It is interesting because superradiant quantum phase transitions that exists here can be controllably implemented~\cite{lab17,lab18}. We analytically investigate the quantum thermometry by using the quantum Langevin equation, standard mean field theory and adiabatic elimination. Then we obtain the quantum Fisher information near the CP, demonstrating that the transition from the normal phase to superradiant phase can be used to improve the measurement precision greatly and in proximity of the CP the direct photon detection is close to the optimal measurement. When the extra thermal bath interacting with the cavity field is independent of the thermal bath to be tested, the measurement precision of the temperature always increases with the coupling strength. When the extra thermal bath interacting with the cavity field is in thermal equilibrium with the thermal bath to be tested, the information of the temperature carried by the interaction is less than the noise it produces until the coupling strength is close to the CP.

This article is organized as follows. In Section II, we simply introduce the quantum metrology and prove that three practical versions of quantum Fisher information with a single-mode Gaussian state are equivalent. In Section III, we elaborate the quantum Rabi model and give the quantum Langevin equation. In Section IV, the
steady-state solutions are achieved by mean field approximation and the system only in the normal phase is stable. In Section V, we investigate the quantum thermometry by adiabatic elimination for two cases. We make a brief conclusion in Section VI.

\section{Review of quantum metrology and three practical versions of quantum Fisher information}
Quantum metrology is a fundamental and important subject concerning the estimation of parameters, including non-unitary parameter (temperature), under the constraints of quantum mechanics~\cite{lab20}. The whole measurement process can be divided into three steps: (1) encoding the unknown parameter into an appropriate probe state, which can be achieved by two ways; one is based on the time-evolution state induced by a parameterized Hamiltonian, the other one is based on the ground state in the Hamiltonian case, or the system steady state when driven-dissipative systems are considered~\cite{lab21,lab22,lab23,lab24}; (2) gathering data by repeated measurements on the probe state; (3) combining the gathered data into an estimator to deduce the value of the parameter~\cite{lab25}. The famous Cram\'{e}r-Rao bound~\cite{lab26,lab27,lab28} offers a good estimation limit of parameter precision

\begin{align}
(\delta \theta)^2\geq\frac{1}{NF[\hat{\rho}(\theta)]},
\label{eq:1}
\end{align}
where $N$ represents total number of repeated experiments. $F[\hat{\rho}(\theta)]$ denotes quantum Fisher information (QFI), which can be generalized from classical Fisher information. The classical Fisher information is defined by
\begin{align}
f(x)=\sum_k p_k(\theta)[d\ln[p_k(\theta)]/d\theta]^2,
\label{eq:2}
\end{align}
where $p_k(\theta)$ is the probability of obtaining the set of experimental results $k$ for the parameter value $\theta$. Furthermore,
the QFI is given by the maximum of the Fisher information over all measurement strategies allowed by quantum physics:
\begin{align}
F[\hat{\rho}(\theta)]=\max_{\{\hat{E}_k\}}f[\hat{\rho}(\theta);\{\hat{E}_k\}],
\label{eq:3}
\end{align}
where positive operator-valued measure $\{\hat{E}_k\}$ represents a specific measurement device.

Gaussian state is a widely-used quantum state in quantum physics, particularly in quantum thermometry. For a single-mode Gaussian state, there are three practical versions of quantum Fisher information.
We introduce a uniform way of definition. The quadrature operators are defined as $q :=\frac{1}{\sqrt{2}}(a+a^\dagger)$ and $p :=\frac{1}{i\sqrt{2}}(a-a^\dagger)$ ($\hbar=1$) with $a$ ($a^\dagger$) as the annihilation (creation) operator for a single bosonic mode (from now on, the hats on the operators are omitted to simplify the description). A vector of quadrature operators is $\mathbf{X}=(q,p)^\top$. The covariance matrix $\mathcal{C}$ with the entries defined as $\mathcal{C}_{ij}:=\frac{1}{2}\langle \{\mathbf{X}_i,\mathbf{X}_j\}\rangle-\langle \mathbf{X}_i\rangle\langle \mathbf{X}_j\rangle$, where $\langle\bullet\rangle=Tr[\bullet\rho_\theta]$. The symplectic matrix is defined as $K:=2i \sigma_y$, where $\sigma_y$ denotes the Pauli operator $(\sigma_y=i(|1\rangle\langle0|-|0\rangle\langle1|)/2)$.

The first version of quantum Fisher information is obtained through the fidelity by Pinel \textit{et al.} in 2013~\cite{lab29},
\begin{align}
F[\hat{\rho}(\theta)]=\frac{1}{2(1+P_\theta^2)}Tr[(\mathcal{C}^{-1}_\theta\mathcal{C}^{'}_\theta)^2]+\frac{2P_\theta'^2}{1-P_\theta^4}\nonumber\\+{\langle\mathbf{X}^\top\rangle}'_\theta\mathcal{C}^{-1}_\theta\langle\mathbf{X}\rangle'_\theta,
\label{eq:4}
\end{align}
where $P_\theta=\frac{1}{2 d}$, $d=\sqrt{Det \mathcal{C}}$ and $A'_\theta$ is the term by term derivative of $A_\theta$ with respect to $\theta$.

The second version was described by ~\cite{lab30},
\begin{align}
F[\hat{\rho}(\theta)]=\frac{2(4d^2-1)}{4d^2+1}Tr[\Omega J'_\theta\Omega \mathcal{C}'_\theta]+{\langle\mathbf{X}^\top\rangle}'_\theta\mathcal{C}^{-1}_\theta\langle\mathbf{X}\rangle'_\theta,
\label{eq:5}
\end{align}
where $J_\theta=\frac{1}{4d^2-1} \mathcal{C}_\theta$.

There is also another widespread version which reads~\cite{lab32},
\begin{align}
F[\hat{\rho}(\theta)]&=
\frac{8}{16d^4-1}\{d^4Tr[(\mathcal{C}^{-1}_\theta\mathcal{C}^{'}_\theta)^2]-\frac{1}{4}Tr[(K\mathcal{C}^{'}_\theta)^2]\}\nonumber\\
&+{\langle\mathbf{X}^\top\rangle}'_\theta\mathcal{C}^{-1}_\theta\langle\mathbf{X}\rangle'_\theta.
\label{eq:6}
\end{align}

The above three practical versions are equivalent, and we give a simple proof in the Appendix.A.

\section{dissipative quantum Rabi system and corresponding quantum Langevin equation}
We consider a two-level atom (spin) interacting with a single cavity field mode according to the common quantum Rabi Hamiltonian:
\begin{align}
H=\omega_0a^\dagger a+\Omega\sigma_z+\lambda(a^\dagger+ a)\sigma_x,
\label{eq:7}
\end{align}
where $\omega_0$ is the frequency of the cavity field; $a$ ($a^\dagger$) is the annihilation (creation) operators of the field satisfying communication relation $[a,a^\dagger]=1$; $\sigma_x$ and $\sigma_z$ are the Pauli operators associated with the spin ($\sigma_z=(|1\rangle\langle1|-|0\rangle\langle0|)/2,\sigma_x=(|1\rangle\langle0|+|0\rangle\langle1|)/2)$, and $\lambda$ is the coupling strength.

The spin directly interacts with the thermal bath with the temperature $T$ to be tested. Simultaneously, the cavity also suffers from the extra thermal bath with temperature $T_c$. The whole dissipative dynamics of the quantum Rabi system is describe by a Markovian master equation
\begin{align}
\frac{\partial \rho}{\partial t}=-i[H,\rho]+\kappa(n_c+1) L[a]\rho+\kappa n_c L[a^\dagger]\rho\nonumber\\
+\Gamma(n+1) L[\sigma^-]\rho+\Gamma n L[\sigma^+]\rho,
\label{eq:8}
\end{align}
where $\kappa$ and $\Gamma$ represent the decay rates of the cavity field and the spin,respectively; $\sigma^+=(\sigma^-)^\dagger=|1\rangle\langle0|$, the Lindblad terms read $L[A]\rho=2A\rho A^\dagger-\{A^\dagger A, \rho\}$, the average thermal photon number is $n_c=\frac{1}{\exp(\omega_0/T_c)-1}$, and $n=\frac{1}{\exp(\Omega/T)-1}$.
The Langevin equations corresponding to Eq.~(\ref{eq:8}) can be expressed as~\cite{lab31}
\begin{align}
\dot{Q}&=-\kappa Q+\omega_0 P+\sqrt{2\kappa}(a_{in}+a_{in}^\dagger),\label{eq:9}\\
\dot{P}&=-\kappa P-\omega_0 Q-2\lambda \sigma_x+\sqrt{2\kappa}i(a_{in}^\dagger-a_{in}),\label{eq:10}\\
\dot{\sigma_x}&=-\Omega \sigma_y-\Gamma\sigma_x-\sqrt{2\Gamma}\sigma_z(\sigma^\dagger_{in}+{\sigma_{in}}),\label{eq:11}\\
\dot{\sigma_y}&=\Omega \sigma_x-\Gamma\sigma_y-\lambda Q \sigma_z+\sqrt{2\Gamma}\sigma_z i(\sigma^\dagger_{in}-{\sigma_{in}}),\label{eq:12}\\
\dot{\sigma_z}&=-(4\Gamma n+2\Gamma)\sigma_z+\lambda Q \sigma_y-\Gamma+\sqrt{2\Gamma}(\sigma^-\sigma^\dagger_{in}+\sigma^+{\sigma_{in}}),
\label{eq:13}
\end{align}
where the quadrature operators are $Q=\sqrt{2}q=a+a^\dagger$ and $P=\sqrt{2}p=i(a^\dagger-a)$, the noise operators $a_{in}$, $a_{in}^\dagger$, $\sigma^+_{in}$ and $\sigma^-_{in}$ from the thermal bath are given by
\begin{align}
\langle a_{in}^\dagger\rangle=\langle a_{in}\rangle=\langle\sigma^\dagger_{in}\rangle=\langle\sigma_{in}\rangle=0\\
\langle a_{in}(t)a_{in}^\dagger(t')\rangle=(1+n_c)\delta(t-t'),\\
\langle a_{in}^\dagger(t')a_{in}(t)\rangle=n_c\delta(t-t')\\
\langle \sigma_{in}(t)\sigma^\dagger_{in}(t')\rangle=(1+n)\delta(t-t'),\\
\langle \sigma^\dagger_{in}(t')\sigma_{in}(t)\rangle=n\delta(t-t').
\label{eq:17}
\end{align}
\section{mean field approximation}
In this section, we can obtain the linear Langevin equation using the mean field approximation by expanding an arbitrary operator $A$ in the form of $A=\langle A\rangle+\delta A$. Specifically, $\sigma_z(\sigma^+_{in}+{\sigma^-_{in}})=\langle\sigma_z\rangle(\sigma^+_{in}+{\sigma^-_{in}})$, where a higher order term $\delta\sigma_z (\sigma^+_{in}+{\sigma^-_{in}})$ is ignored.

 Considering $\langle\dot{A}\rangle=0$ for $A=\{Q,P,\sigma_x,\sigma_y,\sigma_z\}$, we can obtain the steady-state values. One solution is $\langle Q\rangle=\langle P\rangle=\langle\sigma_x\rangle=\langle\sigma_y\rangle=0$, $\langle\sigma_z\rangle=-1/(2+4n)$. This solution is trivial. At this point, the system is said to be in the normal phase. There are two nontrivial solutions corresponding to the superradiant phase, which are described by
\begin{align}
\langle Q\rangle=\mp\frac{\sqrt{2}\Delta}{\lambda\sqrt{\omega_0^2+\kappa^2}},\ \langle P\rangle=\mp\frac{\sqrt{2}\Delta\kappa}{\omega_0\lambda\sqrt{\omega_0^2+\kappa^2}},\nonumber\\
\langle\sigma_x \rangle=\mp\frac{\Delta\sqrt{2(\omega_0^2+\kappa^2)}}{2\lambda^2\omega_0},
\langle\sigma_y \rangle=\pm\frac{\Gamma\Delta\sqrt{2(\omega_0^2+\kappa^2)}}{2\lambda^2\omega_0\Omega},\nonumber\\
\langle\sigma_z \rangle=-\frac{(\omega_0^2+\kappa^2)(\Gamma^2+\Omega^2)(1+2n)}{2\lambda^2\omega_0\Omega},
\label{eq:19}
\end{align}
where $\Delta=\sqrt{\lambda^2\omega_0\Omega-(\omega_0^2+\kappa^2)(\Gamma^2+\Omega^2)(1+2n)}$.
$\lambda=\sqrt{\frac{(\omega_0^2+\kappa^2)(\Gamma^2+\Omega^2)(1+2n)}{\omega_0\Omega}}$ is the boundary condition between the normal phase and the superradiant phase.
When $\lambda<\sqrt{\frac{(\omega_0^2+\kappa^2)(\Gamma^2+\Omega^2)(1+2n)}{\omega_0\Omega}}$, $\Delta$ is imaginary so that nontrivial solutions do not exist. Namely, the system is in the normal phase when $\lambda<\sqrt{\frac{(\omega_0^2+\kappa^2)(\Gamma^2+\Omega^2)(1+2n)}{\omega_0\Omega}}$; the system is in the superradiant phase when $\lambda>\sqrt{\frac{(\omega_0^2+\kappa^2)(\Gamma^2+\Omega^2)(1+2n)}{\omega_0\Omega}}$. However, it is unstable in the superradiant phase as shown in Appendix. B.
Hence, in the next part, we only investigate the steady-state case: the normal phase.
\section{adiabatic elimination}
In this section, we consider two cases: $\Gamma\gg\kappa$ and $\Gamma\ll\kappa$. It allows us to apply an adiabatic elimination.
This will help us to analytically study temperature measurements at phase transition points.
\subsection{The first case: $\Gamma\gg\kappa$ }
When $\Gamma\gg\kappa$, the spin system will reach steady state much faster than the cavity field. Let $\dot{\delta \sigma_x}=0,\ \dot{\delta \sigma_y}=0,\ \dot{\delta \sigma_z}=0$ in Eq.~(\ref{eq:A6}), we can obtain that $\delta \sigma_x=\frac{\Gamma\sigma^+_{in}+\Omega\sigma^-_{in}-\lambda\Omega\delta Q}{2(\Omega^2+\Gamma^2)(1+2n)}$.
In the normal phase, substituting $\delta \sigma_x$ into Eq.~(\ref{eq:A6}) to adiabatically eliminate the mode of spin, we can obtain the
evolution equation of the cavity field

 \[
 \left(
\begin{array}{ll}
\dot{\delta Q}\\
\dot{\delta P}\\
  \end{array}
\right )= M_c\left(
\begin{array}{ll}
\delta Q\\
\delta P\\
  \end{array}
\right ) +\\
 \left(
\begin{array}{ll}
\ \ \ \ A^+_{in}\\
A^-_{in}-f_{in}\\
  \end{array}
\right ) ,\]
where the noise operator is described by $f_{in}=\frac{\Gamma\sigma^+_{in}+\Omega\sigma^-_{in}}{(\Omega^2+\Gamma^2)(1+2n)}$ and the matrix operator is given by
 \[M_c=\left(
\begin{array}{ll}
\ \ -\kappa\ \ \ \ \ \ \ \ \ \ \ \ \ \ \ \ \ \ \ \ \ \ \ \omega_0\\
\frac{\lambda^2\Omega}{(\Gamma^2+\Omega^2)(1+2n)}-\omega_0\ \ \ \ -\kappa\\
  \end{array}
\right ) .\]

The eigenvalues of $M_c$ are $E_{\pm}=-\kappa\pm\sqrt{(\frac{\lambda^2\Omega}{(\Omega^2+\Gamma^2)(1+2n)}-\omega_0)\omega_0}$. The eigenvalue $E_{+}<0$ is the condition for the cavity field to be in the stable normal phase. Therefore,  the detailed formula of the stable normal phase condition is derived
\begin{align}
\lambda<\sqrt{\frac{(\omega_0^2+\kappa^2)(\Gamma^2+\Omega^2)(1+2n)}{\omega_0\Omega}}.
\label{eq:20}
\end{align}
It is also the boundary condition between the normal phase and the superradiant phase. $\lambda=\sqrt{\frac{(\omega_0^2+\kappa^2)(\Gamma^2+\Omega^2)((1+2n)}{\omega_0\Omega}}$ is called the critical point (CP) of the normal-superradiant phase transition.
And the characteristic time $\tau$ for the cavity field to reach the steady state is expressed as
\begin{align}
\tau=\frac{1}{\kappa-\sqrt{(\frac{\lambda^2\Omega}{(\Omega^2+\Gamma^2)(1+2n)}-\omega_0)\omega_0}}.
\label{eq:20}
\end{align}

Besides this type of singularity, there is another singularity in the cavity field.
The evolution equation can also be described in the form of the Schr\"{o}dinger-like equation: $i\frac{d}{dt}(\delta Q,\delta P)^\top=H_{\textmd{eff}}(\delta Q,\delta P)^\top+(A^+_{in},A^-_{in}-f_{in})^\top$, where the effective Hamiltonian $H_{\textmd{eff}}=iM_c$. The corresponding eigenvalues of $H_{\textmd{eff}}$ are $E_{\textmd{eff}}=-i\kappa\pm\sqrt{(\omega_0-\frac{\lambda^2\Omega}{(\Omega^2+\Gamma^2)(1+2n)})\omega_0}$. When $\lambda>\sqrt{\frac{\omega_0(\Gamma^2+\Omega^2)(1+2n)}{\Omega}}$,
the eigenvalues are normally complex, denoting that the cavity field is in the anti-PT-symmetry-broken phase regime. When $\lambda<\sqrt{\frac{\omega_0(\Gamma^2+\Omega^2)(1+2n)}{\Omega}}$, the eigenvalues are purely imaginary, representing that the cavity field is in the anti-$\mathcal{PT}$-symmetry phase regime. The condition $\lambda=\sqrt{\frac{\omega_0(\Gamma^2+\Omega^2)(1+2n)}{\Omega}}$ is defined as the exceptional point (EP) in the anti-PT-symmetry cavity field.
Next, we investigate whether the optimal temperature measurement occurs near the EP or the CP.

Supposing the cavity field has reached the steady state after a long-time evolution,
the solutions of $\delta Q$ and $\delta P$ are
\begin{align}
\delta Q=&\int_0^\infty e^{-\kappa t}\{\cosh\sqrt{\omega_0w}tA^+_{in}\nonumber\\
&+\frac{\sqrt{\omega_0}}{\sqrt{w}}\sinh\sqrt{\omega_0w}t(A^-_{in}-f_{in})\}dt,\label{eq:21}\\
\delta P=&\int_0^\infty e^{-\kappa t}\{\cosh\sqrt{\omega_0w}t(A^-_{in}-f_{in})\nonumber\\
&-\frac{\sqrt{w}}{\sqrt{\omega_0}}\sinh\sqrt{\omega_0w}tA^+_{in}\}dt,
\label{eq:22}
\end{align}
where $w=\frac{\lambda^2\Omega}{(\Omega^2+\Gamma^2)(1+2n)}-\omega_0$.
Then, we can obtain the covariance matrix $\mathcal{C}$ with the entries
\begin{align}
\mathcal{C}_{11}&=\frac{1}{2}\{1+2n_{c}-\frac{\omega_0\lambda^2[\omega_0\Gamma(1+2n)+\kappa\Omega(1+n_c)]}{2\kappa\Delta^2}\},\label{eq:23}\\
\mathcal{C}_{22}&=\frac{1}{2}\{1+2n_{c}+\frac{\lambda^2[\omega_0\Gamma(1+2n)-\kappa\Omega(1+n_c)]}{2\kappa\omega_0(\Omega^2+\Gamma^2)(1+2n)}\nonumber\\
&-\frac{\kappa\lambda^2[\omega_0\Gamma(1+2n)+\kappa\Omega(1+n_c)]}{2\omega_0\Delta^2}\},\label{eq:24}\\
\mathcal{C}_{12}&=-\frac{\lambda^2[\omega_0\Gamma(1+2n)+\kappa\Omega(1+n_c)]}{4\Delta^2}.
\label{eq:25}
\end{align}
For $n=0$ and $n_c=0$,  Eq.~(\ref{eq:23}-\ref{eq:25}) recovers the result as shown in Ref.~\cite{lab32}, which decoupled the spin and field by using Schrieffer-Wolff transformation and projected the spin onto the $|0\rangle\langle0|$ space. The advantage of our method is that we can deal with the general case of non-zero temperature.
\begin{figure*}
\includegraphics[scale=0.75]{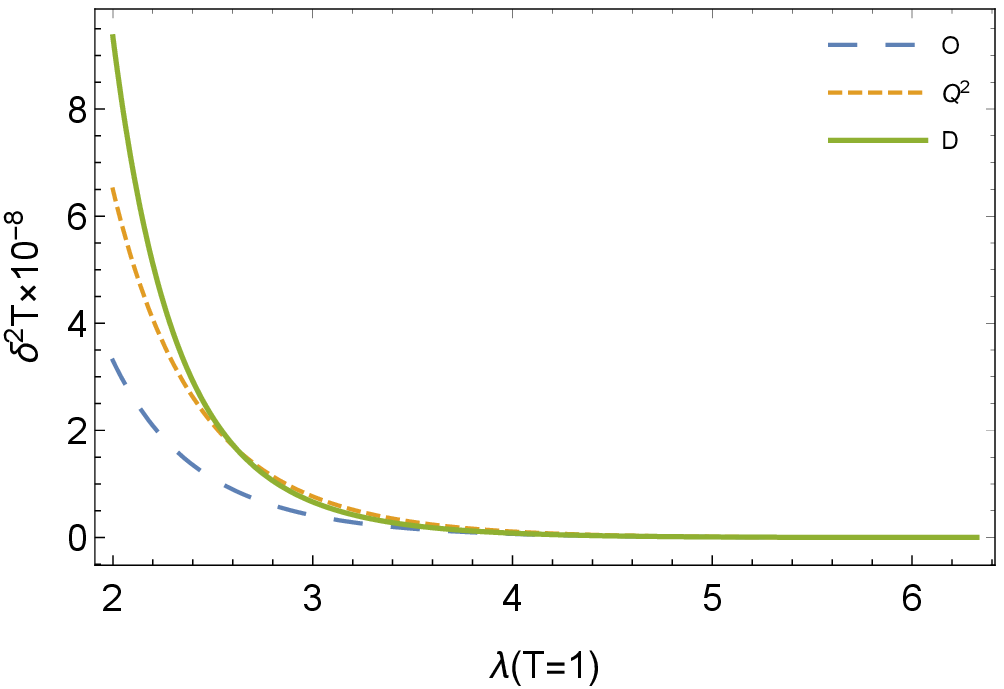}
\includegraphics[scale=0.78]{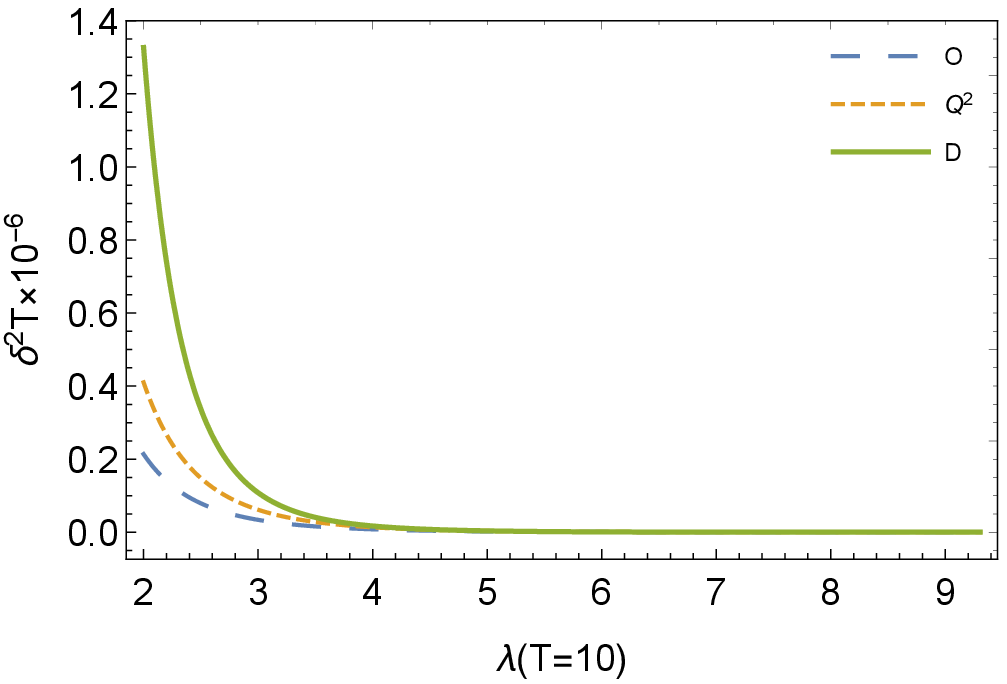}
\includegraphics[scale=0.77]{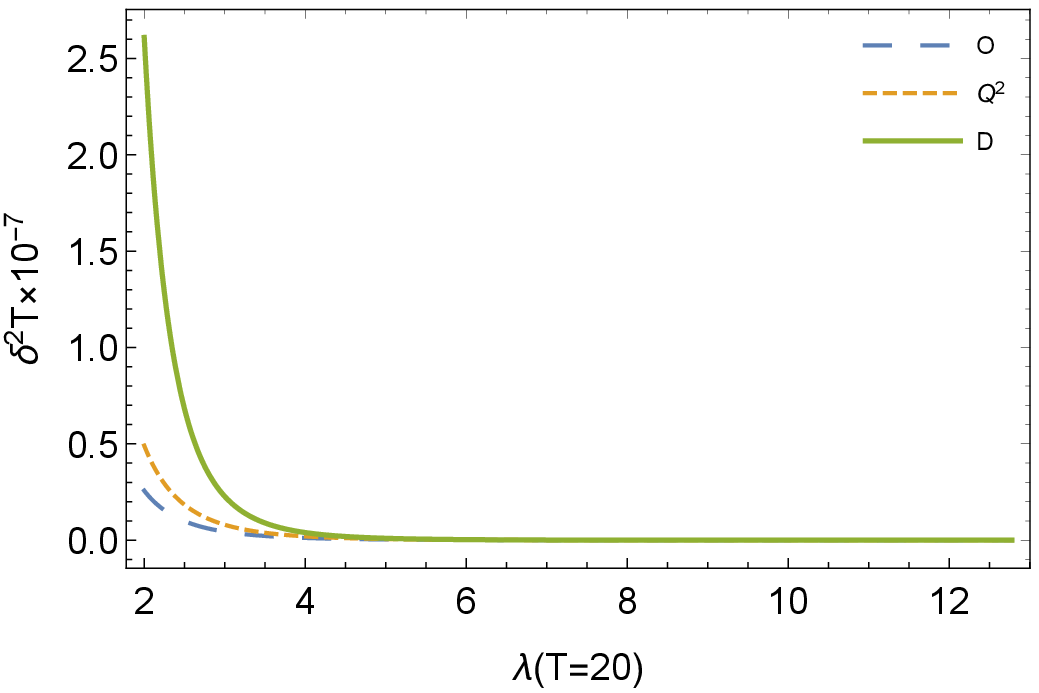}
\includegraphics[scale=0.73]{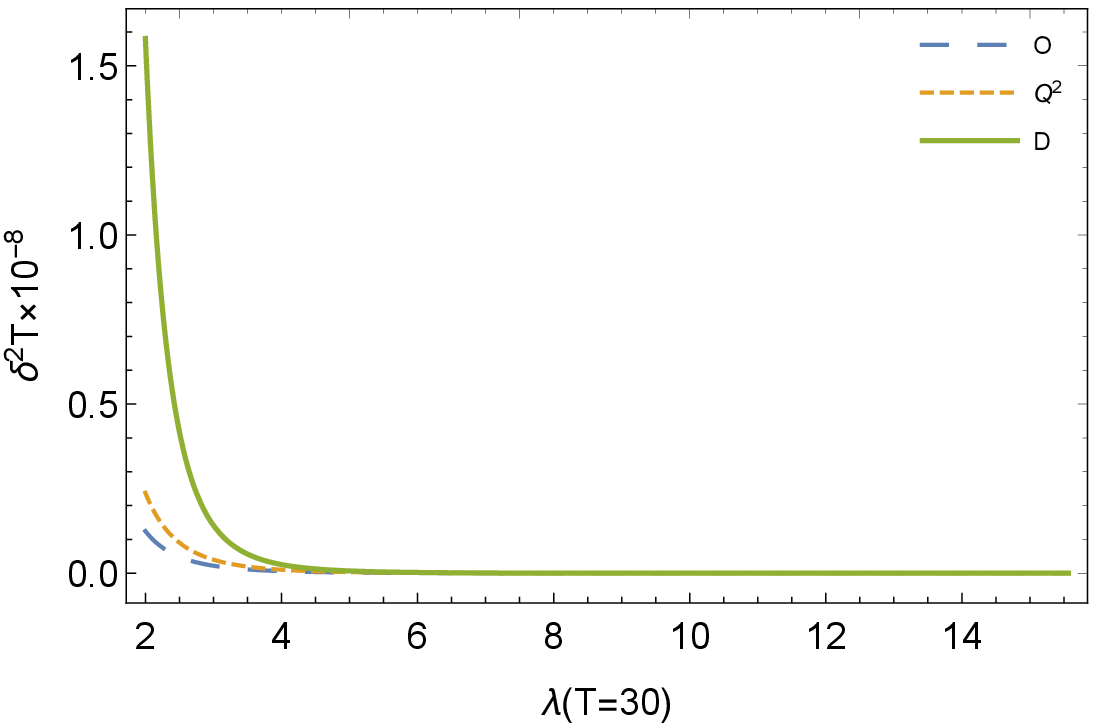}
\caption{\label{fig.1}Graph of temperature measurement precision $\delta^2T$ as a function of the coupling strength $\lambda$ using different measurement methods. The upper limit of $\lambda$ is the CP. The four subgraphs correspond to the thermal bath with temperature $T$ of 1, 10, 20, and 30,respectively. $O$, $Q^2$, and $D$ represent the measurement precision obtained by the quantum Fisher information $F[\hat{\rho}(\theta)]$, the square term of the position operator $Q^2$, and the direct photon detection $a^\dagger a$, respectively. The dimensionless parameters chosen are given by: $N=1$, $T_c=0, \omega_0=1, \Gamma=10, \kappa=1, \Omega=10$.}
\end{figure*}
\begin{figure*}
\includegraphics[scale=0.75]{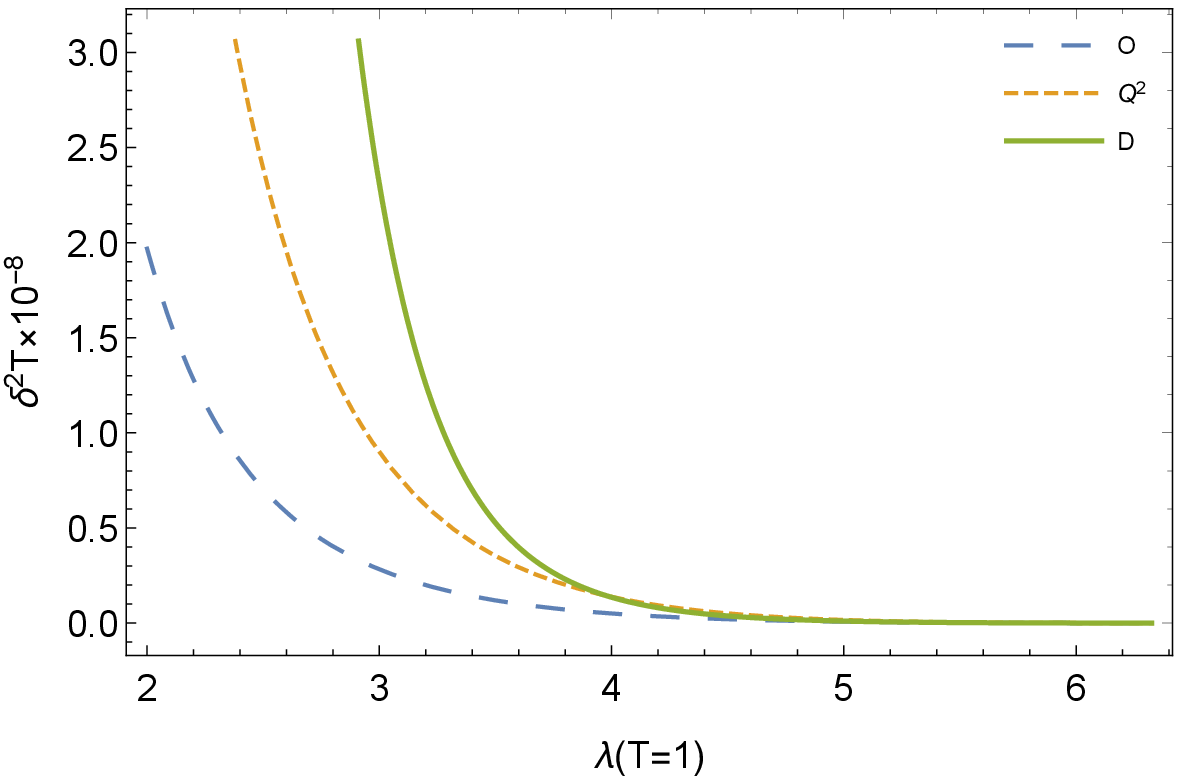}
\includegraphics[scale=0.77]{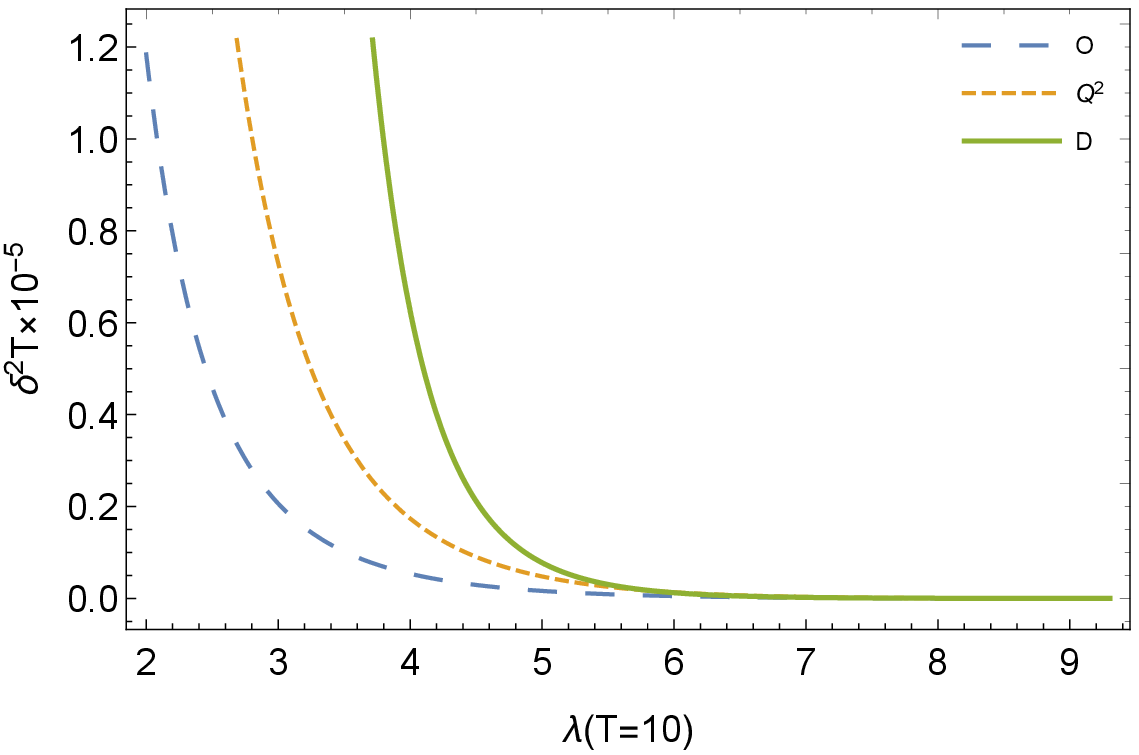}
\includegraphics[scale=0.76]{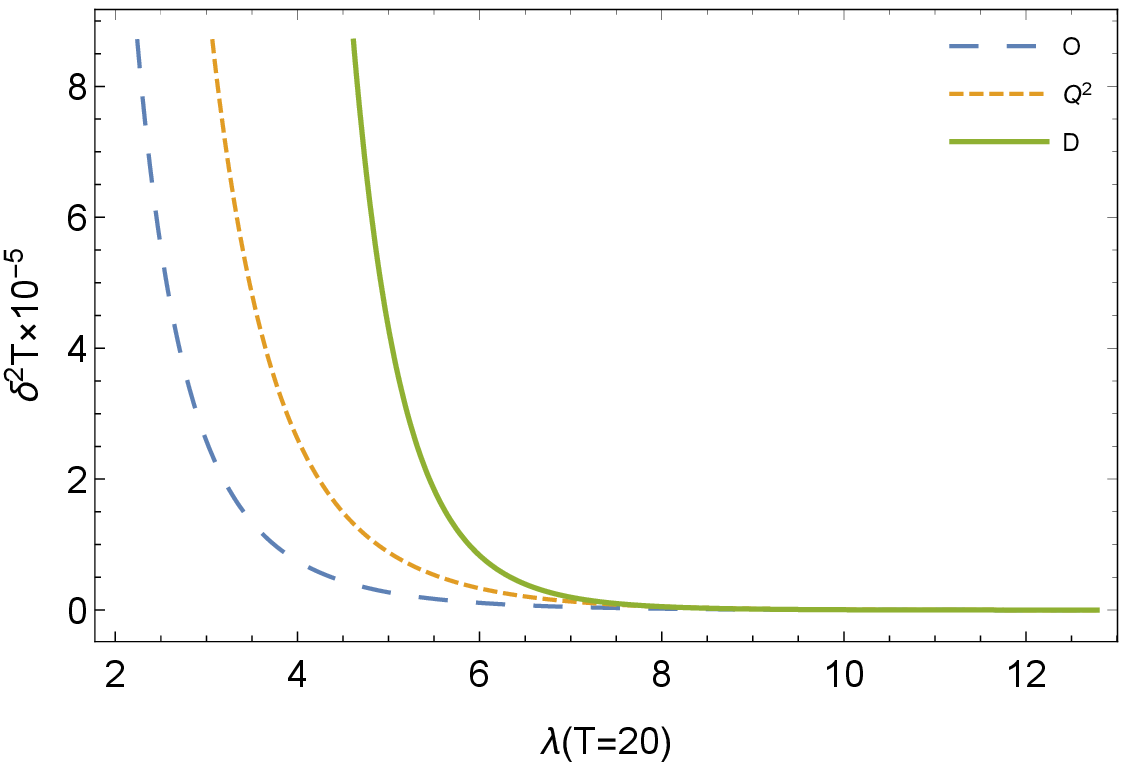}
\includegraphics[scale=0.8]{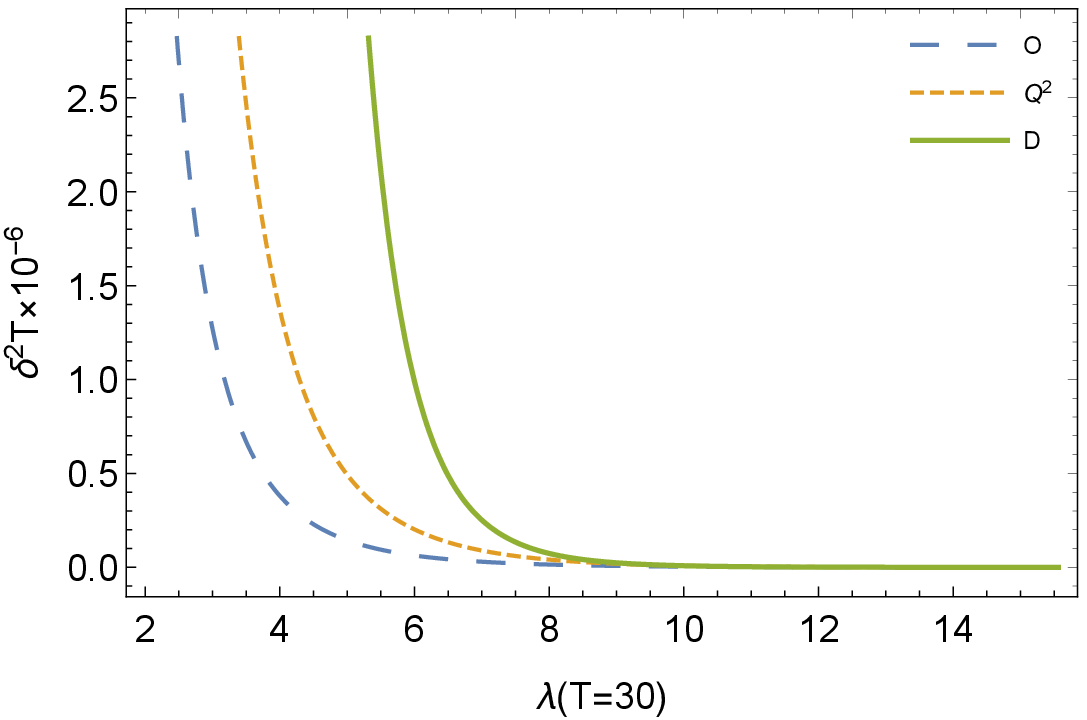}
\caption{\label{fig.2}Graph of temperature measurement precision $\delta^2T$ as a function of the coupling strength $\lambda$ when the temperature of the extra thermal bath interacting with the cavity field is not 0. The upper limit of $\lambda$ is the CP. The four subgraphs correspond to the thermal bath with temperature $T$ of 1, 10, 20, and 30, respectively. The dimensionless parameters chosen are given by: $N=1$, $\omega_0=1, T_c=10, \Gamma=10, \kappa=1, \Omega=10$.}
\end{figure*}

Using any of the three formulas for the quantum Fisher information as shown in Eq.~(\ref{eq:4}-\ref{eq:6}) and the Cram\'{e}r-Rao bound Eq.~(\ref{eq:1}), the optimal estimation precision of the temperature $T$ can be achieved. Near the CP, the leading term of QFI can be achieved

\begin{align}
F[\hat{\rho}(T)]\approx\frac{\tau^2\Omega^2n^2(1+n)^2(\kappa^2+\omega_0^2)^2(\Gamma^2+\Omega^2)^2}{\Delta^4T^4}\label{eq:27}\\
\approx\frac{\Omega^2n^2(1+n)^2(\kappa^2+\omega_0^2)^2}{4T^4(1+2n)^2\kappa^2}
\label{eq:28}
\end{align}

From above equations Eq.~(\ref{eq:27}) and Eq.~(\ref{eq:28}), we can see that as $\lambda$ approaches the CP, the QFI will go to infinity. As a price, the resources of time paid will also tend to infinity. The QFI of the temperature $T$ is proportional to the characteristic time $\tau$ squared. Generally, the measurement precision of temperature is independent of the interaction time, especially in the thermal equilibrium state~\cite{lab321}. It means that the Rabi-interaction makes the information of the temperature be proportional to the characteristic time. This provides a way to greatly improve the precision of temperature measurement.

Because the general and accurate formula of QFI is too long and cumbersome, we carry out numerical processing and then get Fig.~\ref{fig.1} and Fig.~\ref{fig.2}. As a comparison, we calculate the measurement precision of some practical operators by using the error propagation formula, which is described by
\begin{align}
\delta^2T=\frac{\Delta^2O_T}{N|\partial\langle O\rangle_T|^2},
\end{align}
where $\Delta^2O_T=\langle O^2\rangle_T-\langle O\rangle_T^2$ is the variance of the specific operator $O$.
Due to that the first-moment vector is 0, the homodyne detection can not carry the information of the temperature. We mainly consider the feasible measurements ($a^\dagger a$, $Q^2$, and $P^2$). The specific calculation expression is shown in Appendix. C.

\begin{figure*}
\includegraphics[scale=0.85]{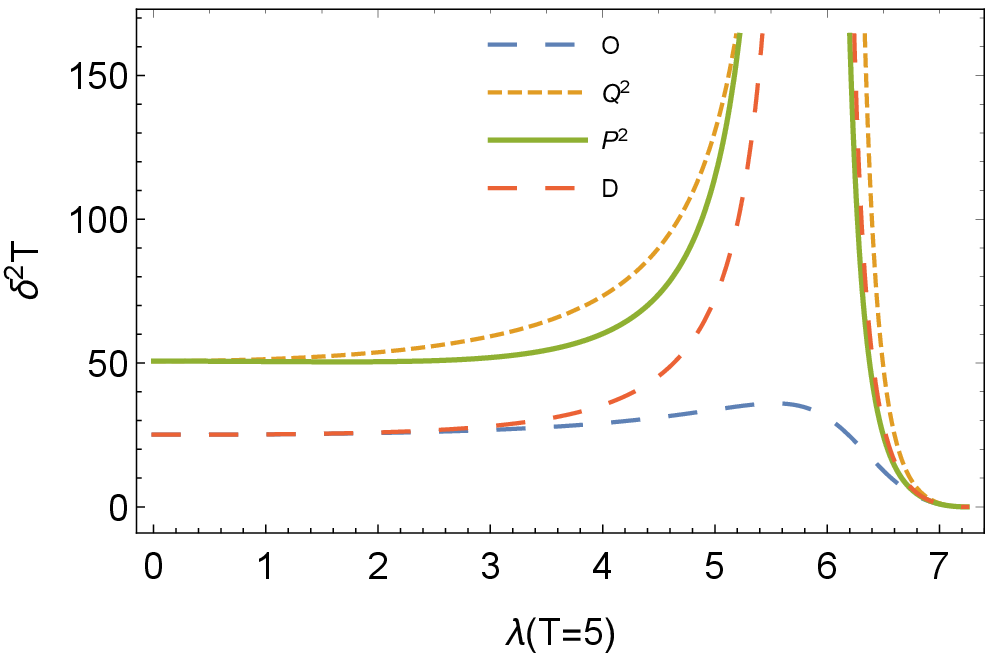}
\includegraphics[scale=0.73]{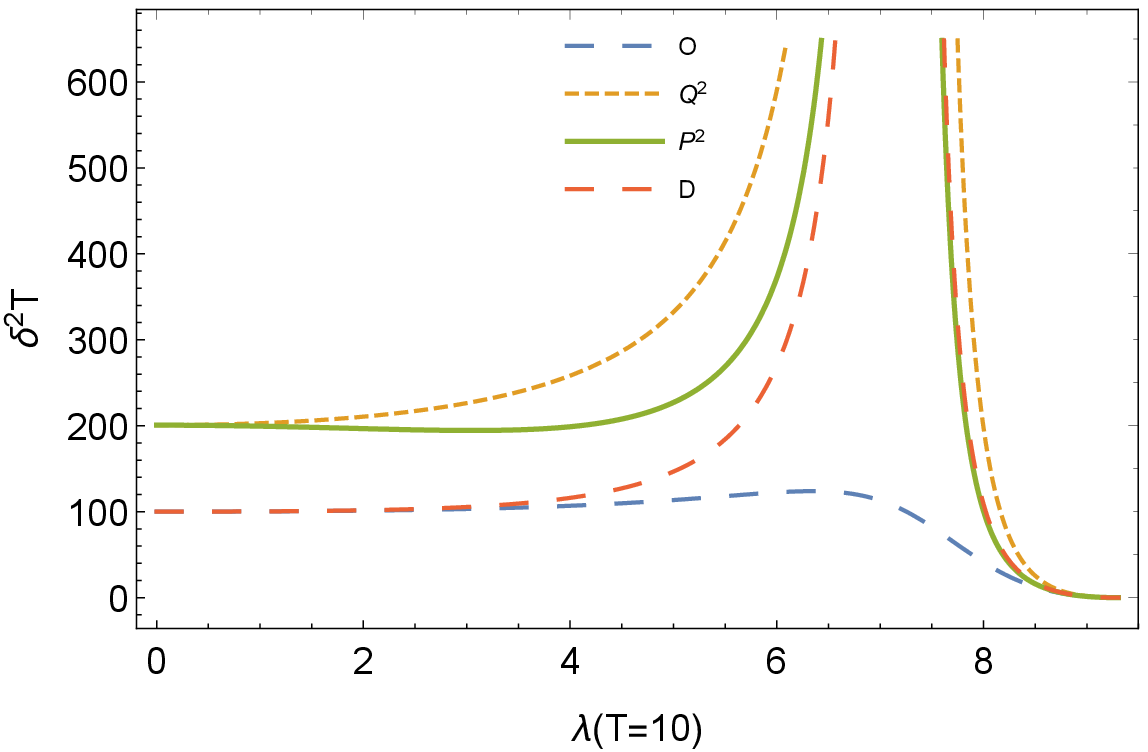}
\caption{\label{fig.3}Graph of temperature measurement precision $\delta^2T$ as a function of the coupling strength $\lambda$ when the extra thermal bath interacting with the cavity field is in thermal equilibrium with the thermal bath to be tested. The two subgraphs correspond to the thermal bath with temperature $T$ of 5 and 10, respectively. The dimensionless parameters chosen are given by: $N=1$, $\omega_0=1,\ \Gamma=10,\ \Omega=10,\ \kappa=1.$}
\end{figure*}
\subsubsection{The extra thermal bath interacting with the cavity field is independent of the thermal bath to be tested}
As shown in Fig.~\ref{fig.1} and Fig.~\ref{fig.2}, we can see that the measurement precision of $T$ is getting higher and higher with the coupling strength $\lambda$. At the CP ($\lambda=\sqrt{\frac{(\omega_0^2+\kappa^2)(\Gamma^2+\Omega^2)(1+2n)}{\omega_0\Omega}}$), the measurement precision of $T$ is optimal. Namely, the optimal measurement precision does not appear at the EP in the anti-$\mathcal{PT}$ symmetrical system. A lot of works~\cite{lab33,lab34,lab35,lab36,lab37} show that the optimal measurement precision can appear around EPs in the $\mathcal{PT}$ symmetrical system. Whether the optimal measurement precision can be found at the EPs of the anti-$\mathcal{PT}$ symmetrical system deserves further rigorous study~\cite{lab38}, which is outside the scope of this article.

For the small coupling strength (away from the CP), the measurement with $Q^2$ performs better than the direct photon detection.  But neither of these is the optimal measurement. Near the CP, the measurement with $Q^2$ and the direct photon detection are all close to the optimal measurement. This finding has nothing to do with the presence or absence of extra thermal baths ($T_c=0$ or $T_c=10$).
\begin{figure}
\includegraphics[scale=1.1]{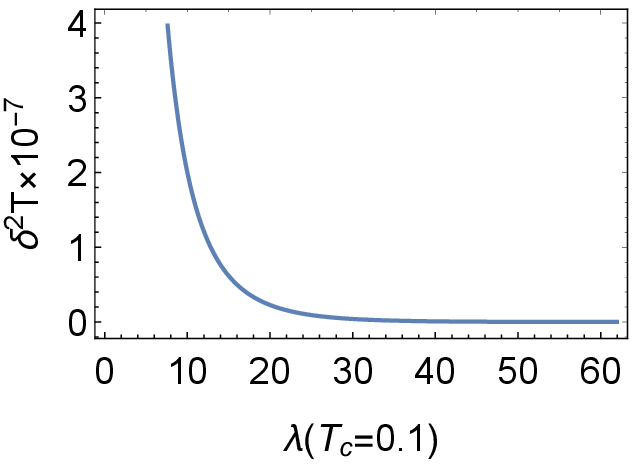}
\includegraphics[scale=1.1]{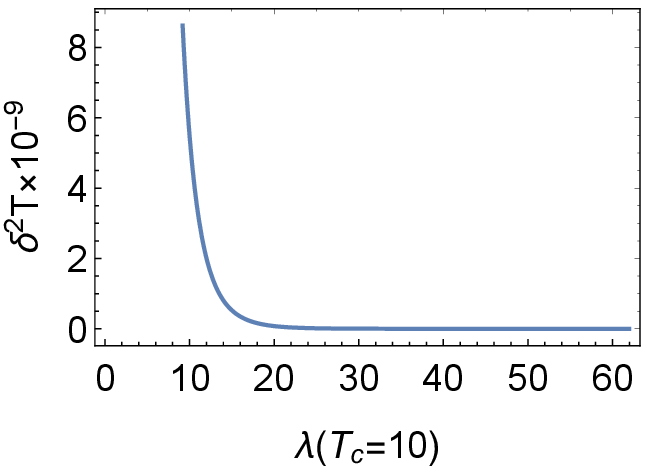}\
{\caption{\label{fig.4}Graph of the optimal temperature measurement precision $\delta^2T$ as a function of the coupling strength $\lambda$ when the cavity field reaches the steady state much faster than the spin system. The two subgraphs correspond to the extra thermal bath with temperature $T_c$ of 0.1 and 10, respectively. The dimensionless parameters chosen are given by: $N=1$, $\omega_0=10,  \Gamma=1, \Omega=1, \kappa=100$.}}
\end{figure}
\begin{figure}
\includegraphics[scale=0.9]{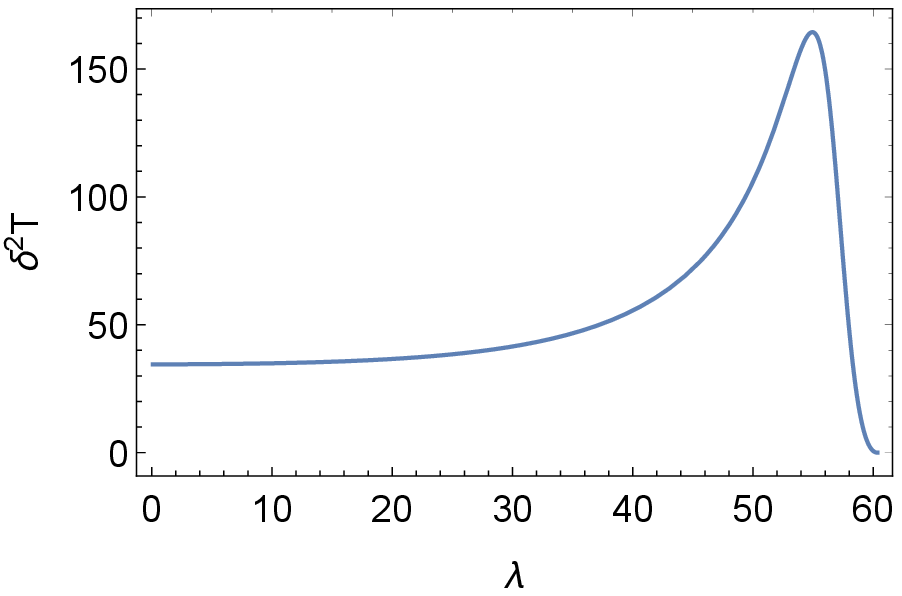}
\caption{\label{fig.5}Graph of temperature measurement precision $\delta^2T$ as a function of the coupling strength $\lambda$ when the extra thermal bath interacting with the cavity field is in thermal equilibrium with the thermal bath to be tested for $\Gamma\ll\kappa$. The dimensionless parameters chosen are given by: $N=1$, $\omega_0=10, T=5, \Gamma=1, \Omega=1,\kappa=100$.}
\end{figure}
\subsubsection{The extra thermal bath interacting with the cavity field is in thermal equilibrium with the thermal bath to be tested}
Due to that the extra thermal bath interacting with the cavity field is in thermal equilibrium with the thermal bath to be tested (for example, interacting with a common thermal bath), the information of $T$ can be encoded into the cavity field by two ways: the first is the interaction between the cavity field and the extra thermal bath; the second is the interaction between the cavity and the spin system, and at the same time the spin system is interacting with the thermal bath to be tested.

As shown in Fig.~\ref{fig.3}, one can find that the direct photon detection performs better than the measurement with operators $Q^2$ and $P^2$, and it is close to the optimal measurement for a small value of $\lambda$. What's more, the optimal measurement precision of $T$ is getting lower and lower with the increase of $\lambda$ when $\lambda$ is smaller than a certain value.
It means that the information of the temperature carried by the Rabi-type interaction is less than the noise it produces in the case of being far away from the CP. It interferes with the temperature information from the first way.

Near the CP, the information of $T$ is mainly obtained by the interaction between the cavity and the spin system. In this case, it carries more information than noise. And the measurement uncertainty obtained by the QFI, $Q^2$, $P^2$ and $a^\dagger a$ get closer and closer to 0 as $\lambda$ gets closer and closer to the CP.
\subsection{The second case: $\Gamma\ll\kappa$ }
When $\Gamma\ll\kappa$, the cavity field will reach steady state much faster than the spin system. With a similar analysis in above subsection, the
evolution equation of the spin system is given by
 \[
 \left(
\begin{array}{ll}
\dot{\delta \sigma_x}\\
\dot{\delta \sigma_y}\\
  \end{array}
\right )= M_s\left(
\begin{array}{ll}
\delta \sigma_x\\
\delta \sigma_y\\
  \end{array}
\right ) +\\
 \left(
\begin{array}{ll}
\ \ \ \ \frac{\sigma^+_{in}}{2(1+2n)}\\
\frac{\sigma^-_{in}}{2(1+2n)}+F_{in}\\
  \end{array}
\right ) ,\]
where the noise operator $F_{in}=\frac{\lambda(\kappa A^+_{in}+\omega_0A^-_{in})}{2(\kappa^2+\omega_0^2)(1+2n)}$ and the matrix is described by
\[M_s=\left(
\begin{array}{ll}
-\Gamma\ \ \ \ \ \ \ \ \ \ \ \ \ \ \ \ \ \ \ \ \ \ -\Omega\\
\Omega-\frac{\lambda^2\omega}{(\kappa^2+\omega^2)(1+2n)}\ \ \ \ -\Gamma\\
  \end{array}
\right ). \]
The eigenvalues of $M_s$ are $E_{\pm}=-\Gamma\pm\sqrt{\frac{\lambda^2\Omega\omega_0}{(\omega^2+\kappa^2)(1+2n)}-\Omega^2}$.
The condition for the stable normal state is derived by $E_{+}<0$. As a result, it is given by
\begin{align}
\lambda<\sqrt{\frac{(\omega_0^2+\kappa^2)(\Gamma^2+\Omega^2)(1+2n)}{\omega_0\Omega}}.
\label{eq:30}
\end{align}
This condition is the same as the steady-state condition in the previous subsection. The CP of the normal-superradiant phase transition is also given by $\lambda=\sqrt{\frac{(\omega_0^2+\kappa^2)(\Gamma^2+\Omega^2)(1+2n)}{\omega_0\Omega}}$.

The analytical formula of the covariance matrix is derived as shown in Appendix. D. When $\lambda$ is close to the CP, the leading term of the QFI is similar with the results as shown in Eq.~(\ref{eq:27}) and Eq.~(\ref{eq:28}).

Numerical results are shown in Fig.~\ref{fig.4} and Fig.~\ref{fig.5}. When the extra thermal bath interacting with the cavity field is independent of the thermal bath to be tested (Fig.~\ref{fig.4}), the measurement precision of the temperature $T$ increases with $\lambda$. And the precision increases very rapidly at small $\lambda$.
When the extra thermal bath interacting with the cavity field is in thermal equilibrium with the thermal bath to be tested (Fig.~\ref{fig.5}), this result is similar to the previous result: the information of $T$ carried by the interaction is less than the noise it produces until $\lambda$ is close to the CP.
\section{conclusion }
We investigate the thermometry of the thermal bath surrounding the spin system with quantum optical probes.
By using the quantum Langevin equation, standard mean field theory and adiabatic elimination, we can recover the results obtained by Schrieffer-Wolff transformation and the projection of spin onto the ground state. Furthermore, our method can deal with non-zero temperature conditions, which highlights the advantages of this approach. We show that the normal state is stable whereas the superradiant state is not. We analytically achieve the QFI near the CP, and show that it is proportional to the square of the characteristic time for arriving at the stable state. This will provide a way to improve the measurement precision of the temperature greatly.
By numerically results, we show that near the CP, the feasible direct photon detection is close to the optimal measurement.
When the extra thermal bath interacting with the cavity field is independent of the thermal bath to be tested, the measurement precision of the temperature always increases with the coupling strength. And the precision increases very rapidly at small $\lambda$. When the extra thermal bath interacting with the cavity field is in thermal equilibrium with the thermal bath to be tested, the information of $T$ carried by the interaction is less than the noise it produces until $\lambda$ is close to the CP.

Our scheme is feasible due to that the quantum Rabi model has been experimentally studied with different
quantum technologies, including circuit-QED~~\cite{lab39}, trapped-ion~\cite{lab40,lab41}, nanoelectromechanical~\cite{lab42},  cavity-QED~\cite{lab43}, and light transport in femtosecond-laser-written waveguide superlattices~\cite{lab44}. Specially, in superconducting circuit QED system, the coupling strength between the transmission
line resonator and the charge qubit can be changed by tuning the magnetic flux~\cite{lab39}.

\section*{Acknowledgements}

We acknowledge Fengxiao Sun for helpful discussion and constructive comments on the manuscript. This research was supported by the National Natural Science Foundation of China under Grant No. 62001134 and Guangxi Natural Science Foundation under Grant No. 2020GXNSFAA159047 and National Key R\&D Program of China under Grant No. 2018YFB1601402-2.

\section*{Appendix. A}
In this section, we prove that Eq.~(\ref{eq:4}) can be derived from Eq.~(\ref{eq:5}) or Eq.~(\ref{eq:6}).
For a single mode Gaussian state, we can obtain
\begin{align}
K \mathcal{C} K=-d^2{\mathcal{C}^{-1}}\tag{A1}.
\label{eq:A1}
\end{align}
Taking the derivative of both sides with respect to $\theta$, we can obtain
\begin{align}
K \mathcal{C}'_\theta K=-d^2({\mathcal{C}^{-1}})'_\theta-2d d'_\theta{\mathcal{C}^{-1}}\nonumber\\
=d^2{\mathcal{C}^{-1}}{\mathcal{C}}'_\theta{\mathcal{C}^{-1}}-2d d'_\theta{\mathcal{C}^{-1}}.\tag{A2}
\label{eq:A2}
\end{align}
Using the above equations and $\textmd{Tr}[\mathcal{C}'_\theta K \mathcal{C} K]=\textmd{Tr}[ K\mathcal{C}'_\theta K \mathcal{C}]$, one can derive that
\begin{align}
\textmd{Tr}[\mathcal{C}'_\theta\mathcal{C}^{-1}]=2d'_\theta/d \tag{A3}.
\label{eq:A3}
\end{align}
Substituting Eq.~(\ref{eq:A1}-\ref{eq:A3}) into Eq.~(\ref{eq:6}), we achieve
\begin{align}
F[\hat{\rho}(\theta)]=
\frac{8}{16d^4-1}\{d^4Tr[(\mathcal{C}^{-1}_\theta\mathcal{C}^{'}_\theta)^2]-\frac{1}{4}Tr[(K\mathcal{C}^{'}_\theta)^2]\}\nonumber\\
+{\langle\mathbf{X}^\top\rangle}'_\theta\mathcal{C}^{-1}_\theta\langle\mathbf{X}\rangle'_\theta\nonumber\\
=\frac{2d^2}{4d^2+1}Tr[(\mathcal{C}^{-1}_\theta\mathcal{C}^{'}_\theta)^2]+\frac{8d_\theta'^2}{16d^4-1}+{\langle\mathbf{X}^\top\rangle}'_\theta\mathcal{C}^{-1}_\theta\langle\mathbf{X}\rangle'_\theta.\tag{A4}
\label{eq:A4}
\end{align}
With the definition $P_\theta=\frac{1}{2 d}$ as shown in Ref.~\cite{lab29}, Eq.~(\ref{eq:4}) can be obtained.

In the same way, substituting Eq.~(\ref{eq:A1}-\ref{eq:A3}) into Eq.~(\ref{eq:5}), we can derive Eq.~(\ref{eq:4}). In Ref.~\cite{lab30}, the formula of quantum information is expressed as
\begin{align}
F[\hat{\rho}(\theta)]=\frac{4d^2-1}{4d^2+1}\textmd{Tr}[K J'_\theta K \mathcal{C}'_\theta]+{\langle\mathbf{X}^\top\rangle}'_\theta\mathcal{C}^{-1}_\theta\langle\mathbf{X}\rangle'_\theta.\tag{A5}
\label{eq:A5}
\end{align}
Comparing with Eq.~(\ref{eq:A5}), it is important to notice that the factor 2 of the first term on the right-hand side is missing in Ref.~\cite{lab30}.
\section*{Appendix. B}
In this section, we analyze the stability of the superradiant phase. The linearized Langevin equation of Eq.~(\ref{eq:9}-\ref{eq:13}) is expressed as
\begin{align}
\dot{h}=\mathbf{M}h+f_{in}.\tag{A6}
\label{eq:A6}
\end{align}
 in which, $h=(\delta Q,\delta P, \delta \sigma_x,\delta \sigma_y,\delta \sigma_z)^\top$, $f_{in}=(A^+_{in},A^-_{in},-\langle\sigma_z\rangle\sigma^+_{in},\langle\sigma_z\rangle\sigma^-_{in},0)$ and the evolution matrix
 \[
 \mathbf{M}= \left(
\begin{array}{ll}
-\kappa\ \ \ \ \ \ \ \omega_0\ \ \ \ \ 0\ \ \ \ \ \ 0\ \ \ \ \ \ \ \ \ 0\\
-\omega_0\ \ \ -\kappa\ \ -2\lambda\ \ \ \ 0\ \ \ \ \  \ \ \ \ 0\\
\ \ 0\ \ \ \ \ \ \ \ 0\ \ \ -\Gamma\ \ \ -\Omega\ \ \ \ \ \ \ 0\\
-\lambda\langle\sigma_z \rangle\ \ \ 0\ \ \ \ \ \Omega\ \ \ \ -\Gamma\ \ \ -\lambda\langle Q\rangle\\
\lambda \langle\sigma_y\rangle\ \ \ \ \ \ 0\ \ \ \ \ 0\ \ \ \ \ \lambda\langle Q\rangle\ \ -(2+4n)\Gamma
  \end{array}
\right ),\tag{A7}
  \]
  where $A^+_{in}=\sqrt{2\kappa}(a_{in}+a^\dagger_{in})$, $A^-_{in}=i\sqrt{2\kappa}(a^\dagger_{in}-a_{in})$, $\sigma^+_{in}=\sqrt{2\Gamma}(\sigma_{in}+\sigma^\dagger_{in})$, and $\sigma^-_{in}=i\sqrt{2\Gamma}(\sigma_{in}-\sigma^\dagger_{in})$.

 The condition for stability is that all the eigenvalues of the matrix $\mathbf{M}$ have
negative real part. Substituting nontrivial solutions in Eq.~(\ref{eq:19}) into the matrix $\mathbf{M}$, there's always an eigenvalue of 0. It means that the superradiant phase is not stable.
\section*{Appendix. C}
By the error propagation formula, the measurement precision by the practical operator can be obtained. In order to deal with the expectation values of the square term of the operator, we use the decoupling relation~\cite{lab50}
\begin{align}
\langle\hat{A}\hat{B}\hat{C}\hat{D}\rangle=\langle\hat{A}\hat{B}\rangle\langle\hat{C}\hat{D}\rangle+\langle\hat{A}\hat{D}\rangle\langle\hat{B}\hat{C}\rangle+\langle\hat{A}\hat{C}\rangle\langle\hat{B}\hat{D}\rangle\nonumber\\
-2\langle\hat{A}\rangle\langle\hat{B}\rangle\langle\hat{C}\rangle\langle\hat{D}\rangle.\tag{A8}
\label{eq:A8}
\end{align}

As a result, the measurement precision of the temperature $T$ from the direct photon detection ($a^\dagger a$), potential energy related term($Q^2$) and kinetic energy related term ($P^2$)  can be expressed as
\begin{align}
\delta^2T|_{a^\dagger a}=\frac{2\mathcal{C}_{11}^2+2\mathcal{C}_{22}^2+4\mathcal{C}_{12}^2-1}{[\partial(\mathcal{C}_{11}+\mathcal{C}_{22})/\partial_T]^2},\tag{A9}\\
\delta^2T|_{Q^2}=\frac{2\mathcal{C}_{11}^2}{[\partial\mathcal{C}_{11}/\partial_T]^2},\tag{A10}\\
\delta^2T|_{P^2}=\frac{2\mathcal{C}_{22}^2}{[\partial\mathcal{C}_{22}/\partial_T]^2}
\tag{A11}
\label{eq:A9}
\end{align}
where $\mathcal{C}_{11}, \mathcal{C}_{22}, \mathcal{C}_{12}$ are the entries of the covariance matrix $\mathcal{C}$ defined in Eq.~(\ref{eq:23}-\ref{eq:25}).
\section*{Appendix. D}
For $\Gamma\ll\kappa$,
 \[
 \left(
\begin{array}{ll}
\dot{\delta \sigma_x}\\
\dot{\delta \sigma_y}\\
  \end{array}
\right )= M_s\left(
\begin{array}{ll}
\delta \sigma_x\\
\delta \sigma_y\\
  \end{array}
\right ) +\\
 \left(
\begin{array}{ll}
\ \ \ \ \frac{\sigma^+_{in}}{2(1+2n)}\\
\frac{\sigma^-_{in}}{2(1+2n)}+F_{in}\\
  \end{array}
\right ) .\tag{A12}\]
The steady-state solutions are derived
\begin{align}
\delta \sigma_x=&\int_0^\infty e^{-\Gamma t}[\cosh(\sqrt{\Omega W}t)\sigma^+_{in}/2\nonumber\\
&-\frac{\sqrt{\Omega}}{\sqrt{W}}\sinh(\sqrt{\Omega W}t)(\sigma^-_{in}/2+F_{in})]dt,\tag{A13}\\
\delta \sigma_y=&\int_0^\infty e^{-\Gamma t}[\cosh(\sqrt{\Omega W}t)(\sigma^-_{in}/2+F_{in})\nonumber\\
&-\frac{\sqrt{W}}{\sqrt{\Omega}}\sinh(\sqrt{\Omega W}t)\sigma^+_{in}/2]dt,\tag{A14}
\label{eq:22}
\end{align}
where $W=\frac{\lambda^2\omega_0}{(\omega^2+\kappa^2)(1+2n)}-\Omega$ and $F_{in}=\frac{\lambda(\kappa A^+_{in}+\omega_0A^-_{in})}{2(\kappa^2+\omega_0^2)(1+2n)}$.

Then, the expectation values of $\delta^2 \sigma_x$ can be achieved

\begin{align}
\langle\delta^2 \sigma_x\rangle=\frac{1}{4}(1+2n)-\frac{\Omega\lambda^2[\Omega\kappa(1+2n_c)+\omega_0\Gamma(1+2n)]}{8\Gamma\Delta^2}
\tag{A15}
\end{align}

For enough time, the steady solutions of $\delta P$ and $\delta Q$ are given  by

\begin{align}
\delta Q&=\frac{-2\lambda\omega_0}{(\omega^2+\kappa^2)(1+2n)}\delta\sigma_x\nonumber\\
&+\int_0^\infty e^{-\kappa t}[\cos(\omega_0 t)A^+_{in}+\sin(\omega_0 t)A^-_{in}]dt\tag{A16}\\
\delta P&=\frac{-2\lambda\kappa}{(\omega^2+\kappa^2)(1+2n)}\delta\sigma_x\nonumber\\
&+\int_0^\infty e^{-\kappa t}[-\sin(\omega_0 t)A^+_{in}+\cos(\omega_0 t)A^-_{in}]dt
\tag{A17}
\end{align}
Then the covariance matrix can be obtained
\begin{align}
&\mathcal{C}_{11}=\frac{2\lambda^2\omega_0^2\langle\delta^2\sigma_x\rangle}{\Lambda^2}+\frac{(1+2n_c)}{2}+2\kappa\omega_0\Omega\lambda^2(1+2n_c)\nonumber\\
&\frac{\kappa^3+2\Gamma\kappa^2+2\Gamma\omega_0^2+\kappa(\omega_0^2+\Gamma^2-W'\Omega)}{\Lambda^2\alpha},\tag{A18}\\
&\mathcal{C}_{22}=\frac{2\lambda^2\kappa^2\langle\delta^2\sigma_x\rangle}{\Lambda^2}+\frac{(1+2n_c)}{2}\nonumber\\
&-\frac{2\omega_0\Omega\kappa^2\lambda^2(\kappa^2+\omega_0^2-\Gamma^2+W'\Omega)(1+2n_c)}{\Lambda^2\alpha},\tag{A19}\\
&\mathcal{C}_{12}=\frac{2\lambda^2\kappa\omega_0\langle\delta^2\sigma_x\rangle}{\Lambda^2}\nonumber\\
&+\frac{\Omega\kappa\lambda^2(\omega_0^2+\kappa^2)(1+2n_c)[(\kappa+\Gamma)^2-\omega_0^2-W'\Omega]}{\Lambda^2\alpha},\tag{A20}
\end{align}
where $\Lambda=(\omega_0^2+\kappa^2)(1+2n)$.


\begin{thebibliography}{9}

\vspace{3mm}
\bibitem{lab1}I. Fr\'{e}rot and T. Roscilde, Quantum Critical Metrology, Phys. Rev. Lett. 121, 020402 (2018).
\bibitem{lab2}K. Macieszczak, M. Gu\c{t}\u{a}, I. Lesanovsky, and J. P. Garrahan, Dynamical phase transitions as a resource for quantum enhanced metrology, Phys. Rev. A 93, 022103 (2016).
\bibitem{lab3}P. Zanardi, M. G. A. Paris, and L. Campos Venuti, Quantum criticality as a resource for quantum estimation, Phys. Rev. A 78, 042105 (2008).
\bibitem{lab4}C. Invernizzi, M. Korbman, L. C. Venuti, and M. G. A. Paris, Optimal quantum estimation in spin systems at
criticality, Phys. Rev. A 78, 042106 (2008).
\bibitem{lab5}D. Schwandt, F. Alet, and S. Capponi, Quantum Monte Carlo Simulations of Fidelity at Magnetic Quantum
Phase Transitions, Phys. Rev. Lett. 103, 170501 (2009).
\bibitem{lab6}T.-L. Wang, L.-N. Wu, W. Yang, G.-R. Jin, N. Lambert, and F. Nori, Quantum Fisher information as a signature of the superradiant quantum phase transition, New J. Phys. 16, 063039 (2014).
\bibitem{lab7}G. Salvatori, A. Mandarino, and M. G. A. Paris, Quantum metrology in Lipkin-Meshkov-Glick critical systems, Phys. Rev. A 90, 022111 (2014).
\bibitem{lab8}S. Greschner, A. K. Kolezhuk, and T. Vekua, Fidelity susceptibility and conductivity of the current in one dimensional lattice models with open or periodic boundary conditions, Phys. Rev. B 88,195101 (2013).
\bibitem{lab9}D. Rossini and E. Vicari, Ground-state fidelity at first-order quantum transitions, Phys. Rev. E 98, 062137 (2018).
\bibitem{lab10}S. S. Mirkhalaf, E. Witkowska, and L. Lepori, Super-sensitive quantum sensor based on criticality in an anti-ferromagnetic spinor condensate, Phys. Rev. A 101, 043609 (2020).
\bibitem{lab11}S.-J. Gu, H.-M. Kwok, W.-Q. Ning, and H.-Q. Lin, Fidelity susceptibility, scaling, and universality in quantum critical phenomena, Phys. Rev. B 77, 245109 (2008).
\bibitem{lab12}S. Fern\'{a}ndez-Lorenzo and D. Porras, Quantum sensing close to a dissipative phase transition: Symmetry breaking and criticality as metrological resources, Phys. Rev. A 96, 013817 (2017).
\bibitem{lab13}S. Wald, S. V. Moreira, and F. L. Semi\~{a}o, In- and out-of-equilibrium quantum metrology with mean-field quantum criticality, Phys. Rev. E 101, 052107 (2020).
\bibitem{lab14}M. M. Rams, P. Sierant, O. Dutta, P. Horodecki, and J. Zakrzewski, At the Limits of Criticality-Based Quantum
Metrology: Apparent Super-Heisenberg Scaling Revisited, Phys. Rev. X 8, 021022 (2018).
\bibitem{lab15}Yaoming Chu, Shaoliang Zhang, Baiyi Yu, and Jianming Cai, Dynamic Framework for Criticality-Enhanced Quantum Sensing,  Phys. Rev. Lett. 126, 010502 (2021).
\bibitem{lab16}Ning Wang, Gang-Qin Liu,1 Weng-Hang Leong, Hualing Zeng, Xi Feng, Si-Hong Li, Florian Dolde,
Helmut Fedder, J\"{o}rg Wrachtrup, Xiao-Dong Cui, Sen Yang, Quan Li, and Ren-Bao Liu, Magnetic Criticality Enhanced Hybrid Nanodiamond Thermometer under Ambient Conditions, Phys. Rev. X 8, 011042 (2018).
\bibitem{lab17}Z. Zhiqiang, C. H. Lee, R. Kumar, K. Arnold, S. J. Masson,
A. Parkins, and M. Barrett, Nonequilibrium phase transition in a spin-1 Dicke model, Optica 4, 424 (2017)
\bibitem{lab18}K. Baumann, C. Guerlin, F. Brennecke, and T. Esslinger, Dicke quantum phase transition with a superfluid gas in an optical cavity, Nature (London) 464, 1301 (2010).
\bibitem{lab20}V. Giovannetti, S. Lloyd, L. Maccone, Quantum-Enhanced measurements: beating the standard quantum limit, Science \textbf{306}, 1330 (2004).
\bibitem{lab21}M. Tsang, Quantum transition-edge detectors, Phys. Rev. A 88, 021801(R)(2013).
\bibitem{lab22}Peter A. Ivanov, Enhanced two-parameter phase-space-displacement estimation close to a dissipative phase transition, Phys. Rev. A 102, 052611 (2020).
\bibitem{lab23}M. Bina, I. Amelio, and M. G. A. Paris, Dicke coupling by feasible local measurements at the superradiant quantum phase transition, Phys. Rev. E 93, 052118 (2016).
\bibitem{lab24}P. A. Ivanov, Steady-state force sensing with single trapped ion, Phys. Scr. 95, 025103 (2020).
\bibitem{lab25}V. Montenegro, U. Mishra, and A. Bayat, Global sensing and its impact for quantum many-body probes with criticality, arxiv: 2102.03843 (2021).
\bibitem{lab26}H. Cram\'{e}r, \textit{Mathematical Methods of Statistics}, (Princeton University, Princeton, 1946).
\bibitem{lab27}C. R. Rao, \textit{Linear Statistical Inference and Its Applications}, (Wiley, NewYork, 1973).
\bibitem{lab28}S. L. Braunstein and C. M. Caves, Statistical distance and the geometry of quantum states, Phys. Rev. Lett. 72, 3439 (1994).
\bibitem{lab29}O. Pinel, P. Jian, N. Treps, C.Fabre, and D. Braun. Quantum parameter estimation using general single-mode Gaussian states, Phys. Rev. A 88, 040102(R) (2013).
\bibitem{lab30}Jing Liu, Haidong Yuan, Xiao-Ming Lu, Xiaoguang Wang, Quantum Fisher information matrix and multiparameter estimation, J. Phys. A: Math. Theor. 53, 023001 (2020).
\bibitem{lab32}Louis Garbe, Matteo Bina, Arne Keller, Matteo G . A. Paris, and Simone Felicetti, Critical Quantum Metrology with a Finite-Component Quantum Phase Transition, Phys. Rev. Lett. 124, 120504 (2020).
\bibitem{lab31}C. Gardiner and P. Zoller, Qauntum Noise: a handbook of Markovian and non-Markovian quantum stochastic methods with applications to quantum optics, Vol. 56 (Springer Science \& Business Media, 2004).
\bibitem{lab321}Luis A. Correa, Mohammad Mehboudi, Gerardo Adesso, and Anna Sanpera, Phys. Rev. Lett. 114, 220405 (2015).
\bibitem{lab33}J. Wiersig, Enhancing the Sensitivity of Frequency and Energy Splitting Detection by Using Exceptional Points:
Application to Microcavity Sensors for Single-Particle Detection, Phys. Rev. Lett. \textbf{112}, 203901 (2014).
\bibitem{lab34}J. Wiersig, Sensors operating at exceptional points: General theory, Phys. Rev. A \textbf{93}, 033809 (2016).
\bibitem{lab35}W. Chen, \c{S}. K. \"{O}zdemir, G. Zhao, J. Wiersig, and L. Yang, Exceptional points enhance sensing in an optical micro-cavity, Nature (London) \textbf{548}, 192 (2017).
\bibitem{lab36} H. Hodaei, A. U. Hassan, S. Wittek, H. Garcia-Gracia, R. El-Ganainy, D. N. Christodoulides, and M. Khajavikhan, Enhanced sensitivity at higher-order exceptional points, Nature (London) \textbf{548}, 187 (2017).
\bibitem{lab37}P.-Y. Chen, M. Sakhdari, M. Hajizadegan, Q. Cui, M. M.-C. Cheng, R. El-Ganainy, and A. Al\`{u}, Generalized parity-time symmetry condition for enhanced sensor telemetry, Nat. Electron. \textbf{1}, 297 (2018).
\bibitem{lab38}C. Chen, L. Jin, and R.-B. Liu, Sensitivity of parameter estimation near the exceptional point of a non-Hermitian system, New Journal of Physics 21, 083002 (2019).
\bibitem{lab39}M. Hofheinz, H. Wang, M. Ansmann, R. C. Bialczak, E. Lucero, M. Neeley, A. D. O \'{C}onnell, D. Sank, J. Wenner,
J. M. Martinis, and A. N. Cleland, Synthesizing arbitrary quantum states in a superconducting resonator, Nature (London) 459, 546 (2009).
\bibitem{lab40}D. Leibfried, R. Blatt, C. Monroe, and D. Wineland, Quantum dynamics of single trapped ions, Rev. Mod. Phys. 75, 281 (2003).
\bibitem{lab41}J. Pedernales, I. Lizuain, S. Felicetti, G. Romero, L. Lamata, and E. Solano, Quantum Rabi Model with Trapped Ions, Sci. Rep. 5, 15472 (2015).
\bibitem{lab42}M. LaHaye, J. Suh, P. Echternach, K. C. Schwab, and M. L. Roukes, Nanomechanical measurements of a superconducting qubit, Nature (London) 459, 960 (2009).
\bibitem{lab43}H. Walther, B. T. Varcoe, B.-G. Englert, and T. Becker, Cavity quantum electrodynamics, Rep. Prog. Phys. 69, 1325 (2006).
 \bibitem{lab44}A. Crespi, S. Longhi, and R. Osellame, Photonic Realization of the Quantum Rabi Model, Phys. Rev. Lett. 108, 163601 (2012).
\bibitem{lab50}J. Naikoo, K. Thapliyal, A. Pathak, S. Banerjee, Probing nonclassicality in an optically driven cavity with two atomic ensembles, Phys. Rev. A 97, 063840 (2018).
\end{thebibliography}
\end{document}